\documentclass[article,twocolumn]{revtex4-1}

\usepackage{amsmath} 
\usepackage{graphicx}
\begin{document}

\title{Bead moving on a uniformly rotating rod studied from an inertial reference frame: Common misconceptions and possible ways to address them}
\author{Toby Joseph}
\email{Electronic mail: toby@goa.bits-pilani.ac.in}
\affiliation{BITS Pilani K K Birla Goa Campus, Goa, 403726, India}
\date{6th August, 2020}

\maketitle

Problems involving rotating systems analyzed from an inertial frame, without invoking fictitious
forces, is something that freshman students find difficult to understand in an introductory mechanics
course. One of the problems that I workout in my intermediate mechanics class (which has
students majoring in physics as well as students majoring in engineering and other science disciplines) 
is that of a bead sliding freely on a rod that is rotating uniformly in a horizontal plane. The motivation
for working out this problem are two fold: (i) to train students in setting up Newton's equations 
of motion by identifying force components and equating to the corresponding mass times accelerations 
and (ii) to make the students familiar with the expressions for acceleration in the polar coordinates that we
introduce at the beginning of the course. After guiding them through the solution which predicts an eventual
radially outward motion for the bead (except for a very special initial condition), a typical question that
gets asked is the following: The math is all fine sir, but how does the bead that is initially at rest at some point
on the rod manage to move radially out if there are no radial forces? This article is about the ways I 
have tried to answer this question over the years and help students understand physics of the
problem better.

\section{Problem statement and solution}
\begin{figure}
\begin{center}
\includegraphics[width=0.6\linewidth]{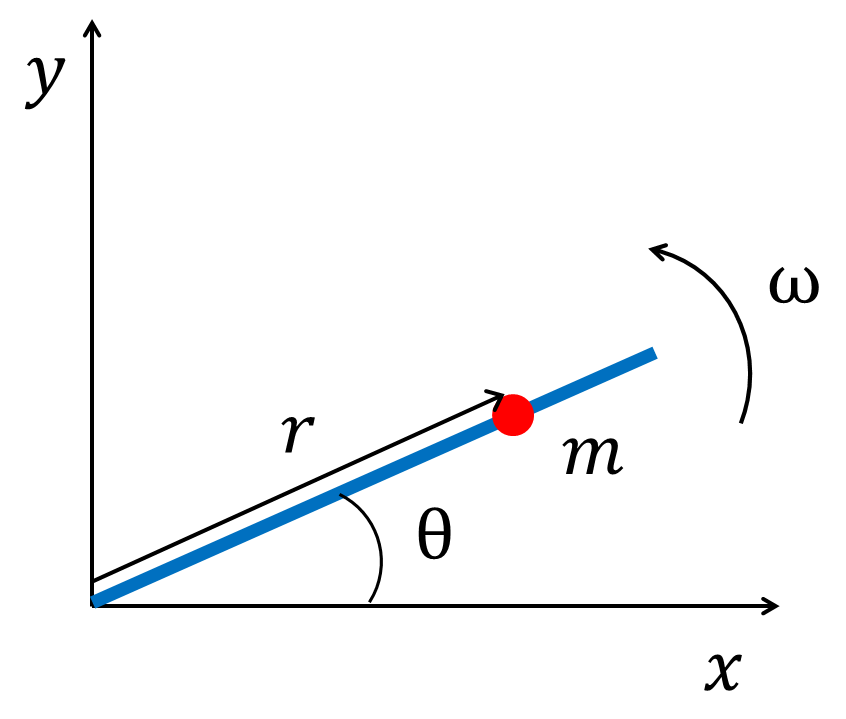} 
\caption{The rod rotates uniformly in a horizontal plane with angular velocity, $\omega$. The bead of mass
$m$ is assumed to slide on the rod without any friction.}
\label{fig1}
\end{center}
\end{figure}

The problem statement is as follows:  A bead of mass $m$ is free to slide on a rod. The rod rotates in the 
horizontal plane about one of its end points with a constant angular speed, $\omega$ (see
Fig.~\ref{fig1}). Find the motion of the bead for arbitrary initial radial velocity and position.

The solution is most easily found by writing down the equations of motion in the polar coordinate system
with origin at the fixed end of the rod (see Fig.~\ref{fig1}). Note that the radial force is always zero, since
the particle is free to slide on the rod. The tangential force is non-zero because the rod can push 
on the bead in the tangential direction. Since we do not know what this force is yet, let us call it $F_{\theta}$.
The radial and tangential equations of motion read
\begin{equation}
m (\ddot r - r \omega^2) = 0
\label{reqn} 
\end{equation}
\begin{equation}
m (r \ddot \theta + 2 \dot r \omega) = F_{\theta}
\label{theqn}
\end{equation}
where we have used the constraint in the problem to put $\dot \theta = \omega$.
The radial equation is readily solved giving
\begin{equation}
r(t) = A e^{\omega t} + B e^{-\omega t}
\label{rsoln}
\end{equation}
where $A$ and $B$ are integration constants to be determined by the initial conditions which are
the radial position and radial velocity at $t = 0$. Note that once the radial solution is found, one
can find the tangential force $F_{\theta}$ as a function of time by using Eq. (\ref{theqn}).

Of particular interest for our discussion here is the case when $\dot r (0) = 0$ and $r(0) = r_0$, 
with $r_0$ not equal to zero. Substituting in Eq. (\ref{rsoln}) and solving we get,
\begin{equation}
r(t) = r_0 \cosh(\omega t) \;.
\label{roft}
\end{equation}
The radial velocity is given by,
\begin{equation}
\dot r(t) = r_0 \omega \sinh(\omega t)
\label{rdt}
\end{equation}
and the tangential force on the bead is
\begin{equation}
F_{\theta} = 2 m r_0 \omega^2 \sinh(\omega t)
\label{tforce}
\end{equation}
which acts in a direction normal to the orientation of the rod at every instant. If we assume that
the rod is aligned along the $x$-axis at $t =0$, then $\theta(t) = \omega t$ and the trajectory
of the bead is give by
\begin{equation}
r(\theta) = r_0 \cosh(\theta) \;.
\label{traj} 
\end{equation}

\section{The confusion and the resolution}
Let us now address the issues that most of the students face with the given approach to the solution.
Some of the key reasons for the difficulty that I could zero in on are listed below. A discussion on how 
I try to address these confusions follows. \\
{\bf (i)} Many students have internalized the fictitious centrifugal force as a `real' force due to their earlier training. 
This is particularly so, since many physics coaching lessons encourage students to use the concept of centrifugal 
force to solve the problem at hand but fail to clarify that one is using a non-inertial frame while doing this 
\cite{doi:10.1119/1.2340592}. \\
{\bf (ii)} Some students still carry the Aristotelian intuition that the force is directly related to velocity rather than
its rate of change. \\
{\bf (iii)} The unit vectors in the polar coordinate system are location dependent and is unlike the 
Cartesian ones they are used to. \\
{\bf (iv)} The more challenging question of understanding how, in the absence of any radial force, 
a particle that started off from rest at some location on the rod can move to a different radial location and 
have a non-zero radial velocity when the rod is back to the same configuration after one rotation.
All of these points or a subset of them might be involved in a students inability to understand the solution
presented.

I use a mix of strategies to address the above points, from pitting their own earlier knowledge against
their faulty intuitions to carrying out calculations to check for consistency of the solution obtained. 
The first point I discuss is the nature of the fictitious forces. To those students who are convinced
about the the reality of the centrifugal force, one can ask them to identify it with one of the four
fundamental forces. Since students are also convinced that there are only four fundamental forces, an
inability to identify the centrifugal force as any of them allows for critical thinking about the nature of
this force. One can then go on to remind them that the centrifugal and Coriolis forces are only relevant
when one uses a non-inertial reference frame which is not what we have done in solving the problem.

At this juncture one can bring up another rotating string problem they are familiar with from high school: the fixed bead 
at the end of a uniformly rotating string. It can be pointed out that in analyzing this problem from the inertial 
frame, centrifugal force need not be (rather, should not be) invoked. It can be stressed that the centripetal 
acceleration is provided by the radially inward force on the bead by the string\cite{doi:10.1119/1.2342581}. 
In fact, it is worthwhile mentioning that even though there is a radially inward force in this case there is no 
radial motion. This is exactly the opposite scenario of the sliding bead on the rod problem where there is 
no radial force but there is non trivial radial motion. Students are thus able to connect the puzzling situation 
in the current problem to a situation they are already familiar with. This helps in reiterating the fact that 
unlike the Aristotelian intuition, the relation between force and motion can be non-trivial.

\subsection{How the radial velocity is generated}
\begin{figure}
\begin{center}
\includegraphics[width=0.6\linewidth]{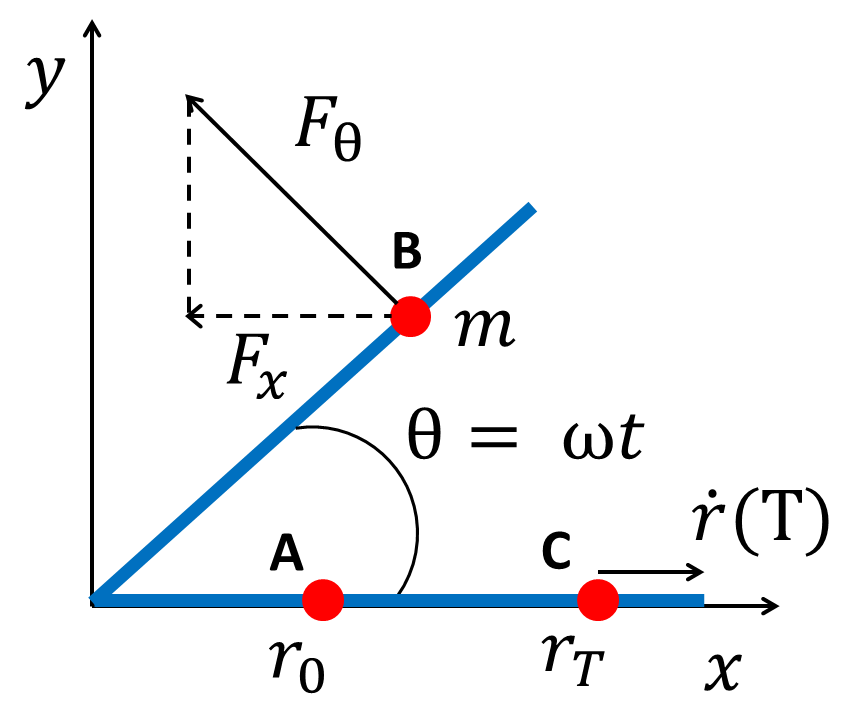} 
\caption{The configuration of the bead and the rod at three different instances: ({\bf A}) at $t = 0$,
({\bf B}) at an intermediate time and ({\bf C}) at $t = T$. The force exerted on the bead by the rod
is $F_{\theta}$ in the tangential direction at every instant. $F_x$ is the projection of this force along
the fixed direction $\hat x$.}
\label{fig2}
\end{center}
\end{figure}

The discussion on the problem is usually wound up by doing a calculation which shows that the observation
in item {\bf (iv)} above is not surprising when one considers the real forces acting on the bead. This calculation 
also brings into focus that the directions of unit vectors in the polar coordinate system are not fixed. Let us 
assume that at $t = 0$, the rod lies along the $x$-axis, and its distance from the fixed end is $r_0$. Additionally,
we shall assume that the radial velocity at $t = 0$ is zero. This means that the solution given in Eq. (\ref{roft})
holds. After one revolution, at $t = \frac{2 \pi}{\omega} \equiv T$, the particle will be at a radial distance
$r(T) = r_0 \cosh(2\pi)$ and it will be moving out with a radial speed
\begin{equation}
\dot r (T) = r_0 \omega \sinh(2 \pi)\;,
\label{rdT}
\end{equation} 
as seen from Eq. (\ref{rdt}) above. We show that the radial velocity that it has picked up during one 
revolution along the $\hat x$ direction is due to the projection of tangential forces that has been acting 
on the bead along the $\hat x$ direction. The initial $(t = 0)$, final $(t = T)$ and an intermediate configurations 
of the rod and the bead are shown in Fig.~\ref{fig2}.
 
The component of tangential force exerted by the rod along $\hat x$ at time $t$ is given by
\begin{equation}
F_x = -F_{\theta} \sin(\omega t)
\label{fx}
\end{equation}
where we have used $\hat \theta = \cos \theta \hat y - \sin \theta \hat x$ (see Fig.~\ref{fig2}).  
We have already determined $F_{\theta}$ (see Eq. (\ref{tforce}) and so are in a position to check for the 
consistency condition that
\begin{equation}
\int_0^T \frac{F_x}{m} dt = \dot r(T)\;.
\end{equation}
Substituting for $F_x$ and carrying out the integration on the LHS, we get
\begin{equation}
\int_0^{\frac{2 \pi}{\omega}}-2 r_0 \omega \sinh(\omega t) \sin (\omega t) dt = r_0 \omega \sinh(2 \pi) \;.
\end{equation}
This is the same as result for $\dot r(T)$ given in Eq. (\ref{rdT}) and thus is consistent. By taking the 
students through this consistency check they are made aware of the fact that the radial velocity has been generated 
by the force that the rod exerts on the bead and not by any mysterious centrifugal force. Though we have
discussed the motion of the particle in the context of a specific problem, the questions we have addressed
come up in a variety of situations involving rotations. For example, similar doubts could crop up in a discussion
of how a centrifuge works \cite{doi:10.1119/1.2344509} or motion of particle sliding in the presence of 
friction \cite{doi:10.1119/1.4896664, doi:10.1119/1.4947157} on a rotating table. A clear understanding of the
simple system we have considered will help students to better grasp the physics of these similar problems.

%

\end{document}